# Three-Dimensional Mid-Infrared Photonics

Recent Progress in Ultrafast Laser Writing of Waveguides

A. Ródenas[1*], R. R. Thomson[1], G. Martin[2], P. Kern[2], and A. K. Kar[1]

[1]School of Engineering and Physical Sciences, Heriot Watt University, Edinburgh EH14 4AS, Scotland, United Kingdom
[2]UJF-Grenoble 1/ CNRS-INSU, Institut de Planétologie et d'Astrophysique de Grenoble (IPAG), UMR 5274, France
[*]a.rodenas@hw.ac.uk

*Abstract*— **We present here our recent progress in the three-dimensional (3D) direct laser writing (DLW) of step-index core waveguides inside diverse technologically relevant dielectric substrates, with specific emphasis on the demonstration of DLW mid-infrared waveguiding in the whole transparency range of these materials.**

*IR photonics; lithium niobate; YAG; chalcogenide; laser writing; laser microfabrication; three-dimensional photonics*

## I. Introduction

The mid-infrared (MIR) range of the electromagnetic spectrum, between 3 μm to 25 μm (100-12 THz), is a key region for a large number of photonic applications such as: sensing in the medical, industrial and environmental fields, high resolution molecular spectroscopy, remote thermal imaging in modern satellites, space science, free-space communication, automotive and aerospace industries, and ground or spaceborne astronomy. MIR light suffers from less Rayleigh scattering than its near-IR or Visible counterparts, and it is also the region where the Earth's atmosphere exhibits important transmission windows for either Earth or space observation: the so-called L (3-4 μm), M (4.6-5 μm) and N (8-12 μm) bands. The MIR is the spectral region which covers the fingerprints of most molecular energy transitions, enabling a vast amount of applications to be developed such as for instance the direct detection of greenhouse gases like $CO_2$ and $CH_4$, or of pollutants such as HCN and $CF_4$. Within astronomy, the MIR is also an exciting observing window, for example to distinguish the heat signature of an Earth-like planet from its host star, to reveal the chemical make-up of remote planets in the search for essential bio-markers such as $H_2O$, $CO_2$ or $O_3$, or to study warm and distant objects, like galaxies, newly forming stars, growing black-holes, comets, asteroids or stellar dust.

Regardless these important potential applications, MIR photonics are still in its maiden infancy. IO technologies are now extremely well developed for the visible and near-infrared range; but the situation is quite different for the MIR spectral domain. Despite a number of two-dimensional (2D) MIR lithographic fabrication approaches have been recently demonstrated, these are all based on multiple-step surface processing and/or make use of thin-films for vertical confinement, therefore constraining the optical designs to strict in-plane propagations.

In this presentation we review our recent progress on the DLW of waveguides (Wgs) for near future 3D MIR photonic technologies. The Wgs are fabricated by an ultrafast ultrashort-pulse laser-writing process which enables the incorporation of the optical circuits into any other instrument device or substrate, as far as the material optical absorption band edge is below the laser wavelength, and without requiring any lithography or clean room facility [1]. With the 3D DLW technique the as-fabricated Wgs typically have average propagation losses in the 0.5 dB/cm range, the Wg cores can be sized and tailored in index contrast to match the specific numerical aperture or modality of connecting fibers, and more importantly, they can be spatially positioned at will inside the sample, therefore giving free access to new refractive-index topologies which were formerly impossible with planar fabrication techniques, and allowing to create photonic devices with unique optical properties which open up the range of imaginable designs, such as photonic lanterns [2] or complex evanescent coupling schemes with before unforeseen properties [3].

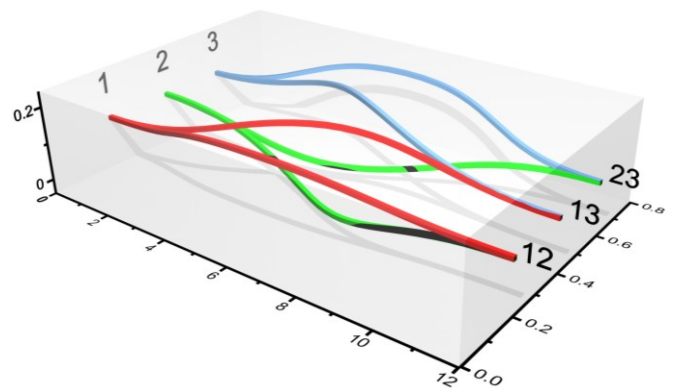

Figure 1. 3D 3-Telescope Beam Combiner for the N-Band. This design was embedded 350 μm deep in the substrate and consists of three input/output combining 3D Y-junctions with zero cross-talk crossovers. All scales are in mm's. [4]

Yet, whereas many of these design possibilities have increasingly been tested and reported since the field of Wg laser writing started in 1996, this has been only done for the visible and near-IR range, and there would be no report of guiding above ~2 μm wavelengths until 2011, when the first Wgs capable of guiding in the whole transparency range of materials such as chalcogenide sulphide glass (≤ 11 μm) were reported [4]. Here we present recent results on the fabrication and preliminary optical characterization of MIR step-index buried Wgs in various transparent substrates such as rare-earth ion doped and undoped lithium niobate (LiNbO$_3$), yttrium aluminum garnet (YAG) [5] and oxoborate (YCOB) crystals [6], or chalcogenide sulphide glass (ChC), and show recent developed devices such as novel beam combiners for astrophotonics applications [4].

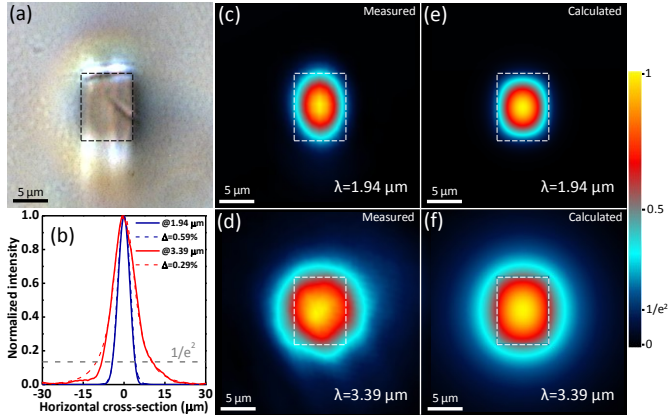

Figure 2. Optical near-IR characterization of a multiscan NdYCOB step-index Wg. (a) Wg core with size of 7.4x9.6 µm$^2$. (b) Horizontal intensity cross-sections of the near-field profiles of fundamental modes at 1.94 µm and 3.39 µm wavelengths. Both measured and calculated modes are shown. (c) and (d) Measured near-field images of single-modes. The waveguide core is also shown. (e) and (f) Corresponding calculated modes and waveguide core.

## II. FEMTOSECOND LASER MICROFABRICATION

For Wg DLW, a 1047 nm wavelength, circularly polarized sub-picosecond pulse laser is used. Initial experiments must be always performed in order to investigate the effects that different pulse characteristics (repetition rate, pulse energy, focusing optics, scan speed, etc) have in the material. Typical pulse repetition rates range between 0.1 to 1 MHz, with pulse duration in the range 350-460 fs. Writing scan speeds usually range between 0.25 to 40 mms$^{-1}$, and laser pulse energies vary from 5 to 300 nJ inside the sample depending on the type of processing that is required. Focusing lenses with numerical apertures (NAs) from 0.25 to 1.4NA are typically used. If the Wg cores need are to be precisely controlled in shape and index contrast, multiple experiments need to be performed before the precise fabrication conditions are found for each type of material, once the writing conditions are determined a precise and repeatable lab manufacturing protocol can be set. DLW step-index waveguides in crystals have proven to be very difficult to achieve, and until recently they had only been demonstrated in LiNbO$_3$ and ZnSe crystals with maximum guiding wavelengths of 1.55 µm and estimated index-contrasts ($\Delta n_{change}/n_o$) lower than 0.14%. We have recently developed novel step-index Wgs in borate, YAG and LiNbO$_3$ crystals with an almost 5-fold increased index-contrast, which allows fabricating Wgs with sufficiently low bend losses for mm size devices to be designed (see Fig.1), as well as for sustaining MIR light confinement in micron size Wg cores.

### A. Borate nonlinear crystals

Fig. 2 shows Wgs performed on a Nd:YCOB nonlinear crystal. A refractive-index change of around $\Delta n \sim 5 \times 10^{-3}$ was estimated from the mode numerical simulation of experimental near field modes at 3.39 µm. This $\Delta n$ yields a core/cladding contrast of $\Delta \sim 0.29\%$ at this wavelength. At 1.94 µm an increased contrast of $\Delta \sim 0.59\%$ was similarly found. This indicates that the contrast at the near-IR (1.55 µm) could also be higher of around $\Delta \sim 0.67\%$.

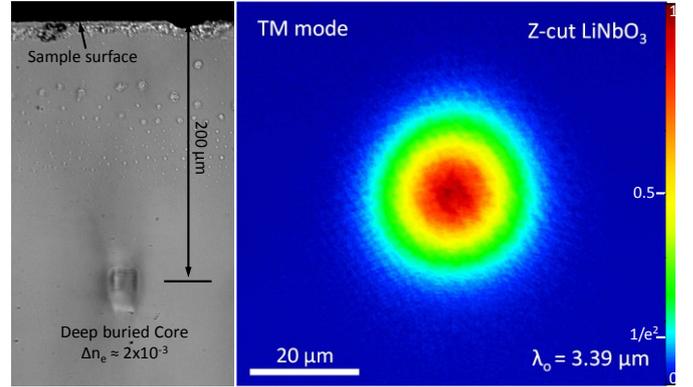

Figure 3. On the left a transmission visible light micrograph of the Wg and its depth in the substrate is shown. On the right, a representative mode of a TM (vertically polarized) single-mode core for 3.39 µm wavelength is shown. MFD at 1/e$^2$ is ~27 µm.

### B. Lithium Niobate

Step-index buried Wgs were also be fabricated in z-cut Er:LiNbO3 and undoped LiNbO3 samples at around 200 µm depth from the top surface (see Fig.3), showing almost single-mode behavior at 3.39 µm for both compositions. These Wgs are eventual building blocks for further developments towards optoelectronic on-chip instruments such as novel 3D Match-Zender interferometers for on-chip fringe scanning, and/or for developing frequency converters elements based on PPLN technology and nonlinear frequency conversion processes for many different applications.

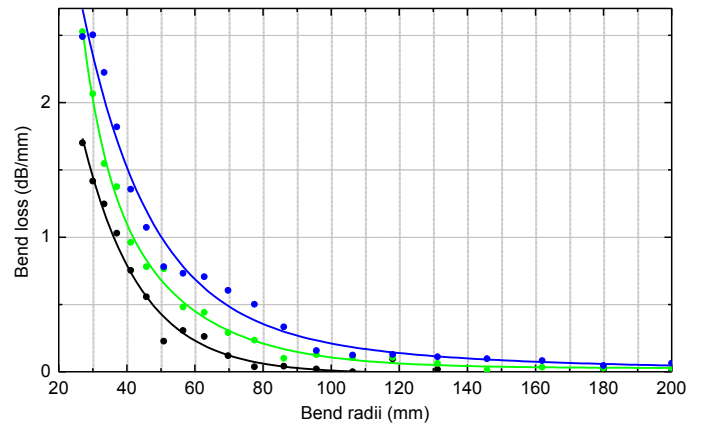

Figure 4. Bend losses as a function of bend radii for step-index 3.39 µm Wgs fabricated with different laser powers of 340 mW (black), 290 mW (green) and 265 mW (blue).

These Wgs could also be written on pre-fabricated chips so that DLW could be used as a complementary fabrication step on hybrid LiNbO3 based on-chip instruments, or, due to their embedded 3D positioning, further substrate processing could be performed once the Wg circuit has been pre-imprinted. In order to evaluate the losses of bent Wgs a preliminary bend loss characterization has been performed shown in Fig.4.

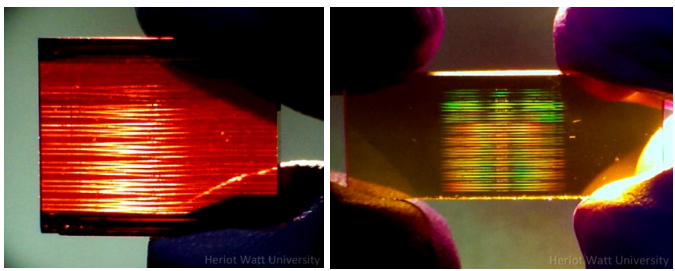

Figure 5. First 3D 3-beam prototype combiners performed in different compositions of chalcogenide sulphide glasses for operation at MIR wavelengths around 11 µm. Their combining design and dimensions are shown in Fig.1. Each sample has 10 different combiners with same combining design but different Wg core characteristics such as core size and index change.

*C. Chalcogenide sulphide glass*

We have investigated commercial and research chalcogenide sulphide glasses as material substrates, both of which are free of the highly toxic arsenic compounds which are typically found in chalcogenide photonics. We have demonstrated the inscription of waveguides which support 3D transitions, enabling the fabrication of arbitrary optical circuits which can also be designed to operate single or multimode throughout the whole transparency window of the chalcogenide glasses (from ~1 µm to ~11 µm), for the first time. To analyze the potential of this technology, we fabricated various novel 3D MIR photonic 3-beam combiners for the N-band. Measurements in the N-band were performed in collaboration with P. Kern astrophotonics group (UJF-Grenoble 1/CNRS-INSU (IPAG) in France.

A characterization of the refractive-index induced changes at 10.6 µm wavelength yielded that the Δn values are typically around $1.2 \times 10^{-2}$. Figure 6 (a) shows the corresponding Δn values as a function of laser parameters as well as the corresponding horizontal MFD of each Wg. Fig. 6 (b) shows an example of a 6 Wg 2D array which was set for calibration. These studies allow us to design future integrated on-chip instruments for the MIR range with more complex designs and functions, as we have demonstrated that it is possible to harness the whole MIR transparency range offered by chalcogenide sulphide glasses on a single on-chip photonic instrument, fabricated by means of 3D DLW. We believe that these findings therefore clearly open a novel technological avenue to new science in a very wide range of mid-infrared applications.

ACKNOWLEDGMENTS

A. Ródenas acknowledges financial support from the Spanish Ministerio de Educación under the Programa de Movilidad de Recursos Humanos del Plan Nacional de I+D+I 2008/2011 for abroad postdoctoral researchers. R. R. Thomson acknowledges support through an STFC Advanced Fellowship (ST/H005595/1). This work was also funded by the UK EPSRC grants EP/F067690/1, EP/G030227/1 and EP/D048672/1.

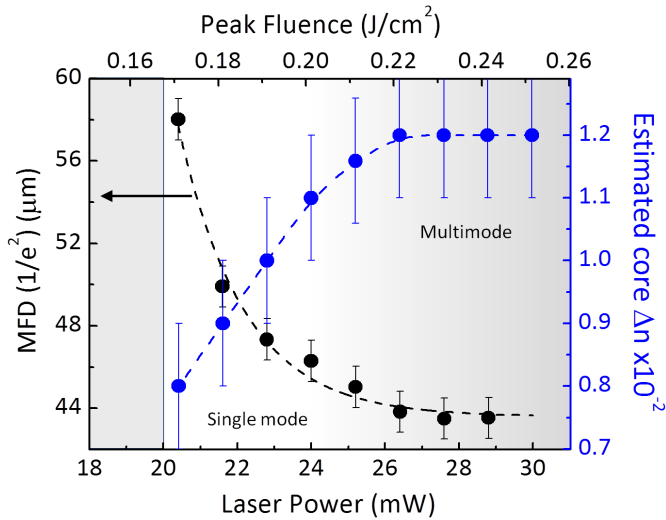

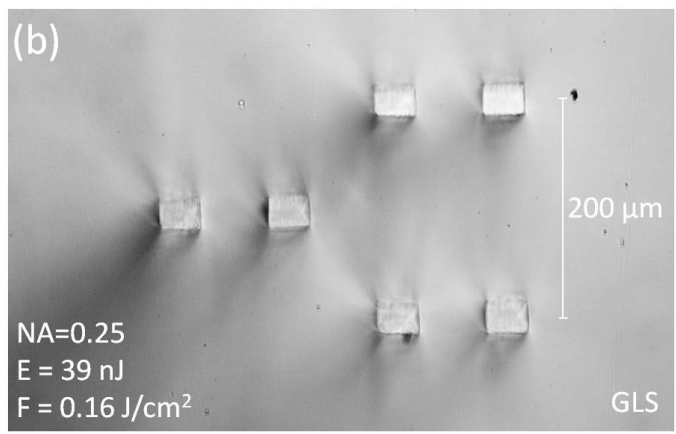

Figure 6. (a) Δn values as a function of laser parameters as well as the corresponding horizontal MFD of each Wg; (b) example of a 6 Wg 2D array which was set for calibration.